# Cavity-enhanced single photon emission from a single impurity-bound exciton

Yuxi Jiang,[1,2] Robert M. Pettit,[1] Nils von den Driesch,[4,5] Alexander Pawlis,[3,4,5] and Edo Waks[1,2]

[1]Institute for Research in Electronics and Applied Physics and Joint Quantum Institute, University of Maryland, College Park, Maryland 20742, USA
[2]Department of Electrical and Computer Engineering, University of Maryland, College Park, Maryland 20740, USA
[3]Peter-Grünberg-Institute (PGI-9), Forschungszentrum Jülich GmbH, 52425 Jülich, Germany
[4]Peter-Grünberg-Institute (PGI-10), Forschungszentrum Jülich GmbH, 52425 Jülich, Germany
[5]JARA-FIT (Fundamentals of Future Information Technology), Jülich-Aachen Research Alliance, 52062 Aachen, Germany

**Abstract**

Impurity-bound excitons in ZnSe quantum wells are bright single photon emitters—a crucial element in photonics-based quantum technology. But to achieve the efficiencies required for practical applications, these emitters must be integrated into optical cavities that enhance their radiative properties and far-field emission pattern. In this work, we demonstrate cavity-enhanced emission from a single impurity-bound exciton in a ZnSe quantum well. We utilize a bullseye cavity structure optimized to feature a small mode volume and a nearly Gaussian far-field transverse mode that can efficiently couple to an optical fiber. The fabricated device displays emission that is more than an order of magnitude brighter than bulk impurity-bound exciton emitters in the ZnSe quantum well, as-well-as clear anti-bunching, which verifies the single photon emission from the source. Time-resolved photoluminescence spectroscopy reveals a Purcell-enhanced radiative decay process with a Purcell factor of 1.43. This work paves the way towards high efficiency spin-photon interfaces using an impurity-doped II-VI semiconductor coupled to nanophotonics.

**Introduction**

Single photon sources are essential components of photonics-based quantum information processors[1–3], quantum networks[4–6], and photonic quantum simulators[7–11]. Among the various approaches of generating single photons, impurity-bound excitons in II-VI wide-bandgap semiconductors have emerged as a promising material platform for realizing efficient single photon emission[12–17]. These emitters also possess a natural spin ground state that can act as a qubit, opening the possibility to realize efficient spin-photon interfaces[12,15,17,18]. In addition, ZnSe quantum wells can be grown isotopically purified[19] to deplete the nuclear spin background, which should lead to a drastic increase of the spin dephasing times. However, fully harnessing the potential of these emitters for quantum applications requires methods to extract their photon emission with high efficiency.

Optical cavities offer the potential to improve both the collection efficiency and radiative properties of impurity-bound exciton emission. These cavities can redirect the emission efficiently out of the substrate as well as shape the transverse mode to better align with the collection optics[20,21]. Additionally, optical cavities can enhance the emitter brightness through the Purcell effect[22–25]. However, coupling these single quantum emitters in II-VI materials to cavities is difficult. First, nanophotonic cavities typically require an epitaxial layered structure featuring a sacrificial layer to form suspended membranes to improve the confinement of light in the optical device, which is challenging to fabricate in II-VI materials. Additionally, II-VI wide-bandgap semiconductors generally emit at the blue and UV wavelengths, thus requiring small cavity feature sizes that increase fabrication difficulty.

In this letter, we demonstrate cavity-enhanced single-photon emission from an impurity-bound exciton emitter in a II-VI semiconductor. The emitter is composed of a Cl impurity in a ZnSe quantum well[12]. To achieve the appropriate layered structure for nanophotonic cavity integration, we grow the II-VI material stack on top of an AlGaAs sacrificial layer, which can be selectively wet etched. We subsequently pattern a nanophotonic bullseye cavity that exhibits both a small mode volume and a Gaussian-like transverse mode that can be efficiently collected through a free-space objective lens to an optical fiber. By coupling a single Cl-bound exciton to this cavity, we achieve an average 15.8-times improvement in brightness relative to emitters in the bulk. Additionally, using time-resolved photoluminescence measurements, we observe a Purcell enhancement of 1.43. The significant enhancement of the brightness and the modification of the radiative decay rate represent an important step towards cavity integration of II-VI impurity-bound exciton emitters for efficient single photon generation and strong spin-photon interfaces.

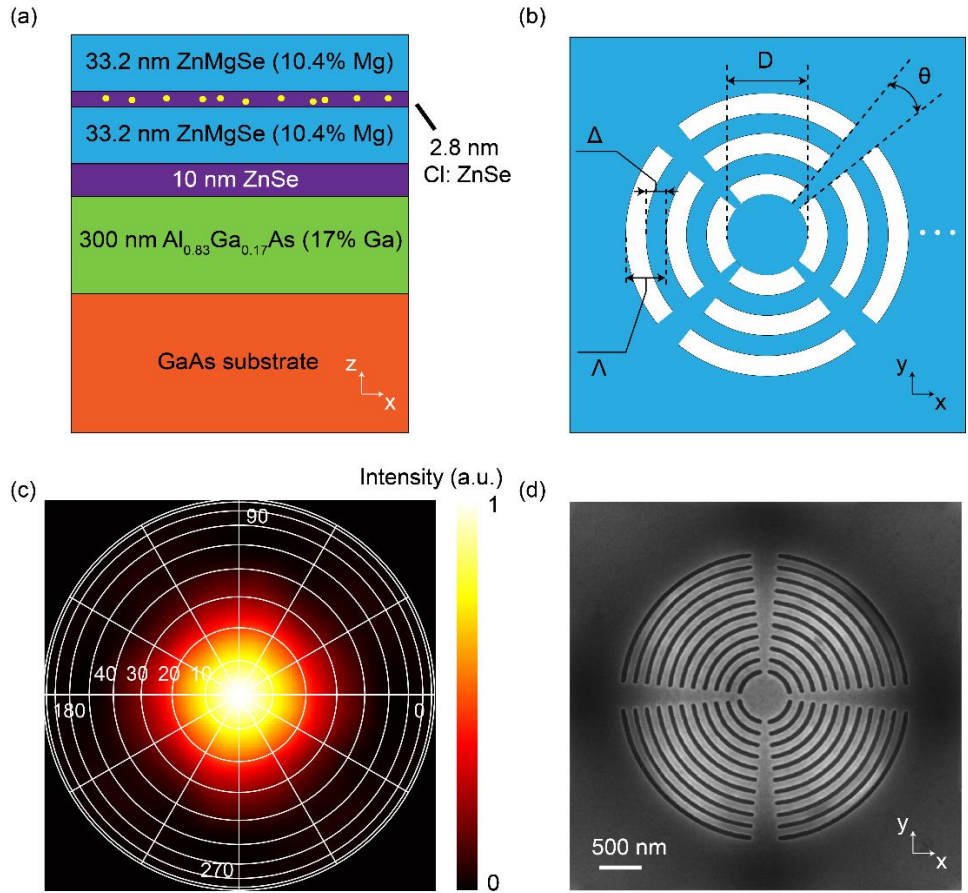

Figure 1. (a) Layered structure of the ZnSe quantum well with delta-doped Cl impurities. (b) A schematic drawing of the bullseye optical cavity on ZnSe quantum well membranes. The white region shows the etched area. (c) Finite-difference time-domain simulation of the far-field radiation profile of a dipole source in the optimized bullseye optical cavity. The simulated profile exhibits a nearly Gaussian distribution. (d) A scanning electron microscopy image showing the fabricated device.

Figure 1a shows the layer structure of the initial wafer, grown by molecular beam epitaxy. The quantum well structure is formed by growing a ZnSe quantum well with layer thickness of 2.8 nm embedded in ZnMgSe cladding layers (10.4% Mg) with layer thicknesses of 33.2 nm. Cl donors are delta-doped into the ZnSe layer at a low concentration density of $10^{10}$ cm$^{-2}$ to achieve spatially and spectrally isolated emitters. The quantum well structure is grow on an Al$_x$Ga$_{(1-x)}$As sacrificial layer with 17% Ga (x = 0.83), with an intermediate ZnSe buffer layer of about 10 nm thickness to facilitate lattice matching. A photoluminescence spectrum from the grown sample is provided in Section I of the supplementary material.

To enhance emission from impurity-bound excitons, we employed the bullseye cavity structure illustrated in Figure 1b. This structure consists of a central disk with diameter D, surrounded by periodic concentric rings with periodicity $\Lambda$ and width $\Delta$. The rings act as Bragg-mirror that confines light in the in-plane direction[26–28]. The cavity structure is suspended by four bridges, each of which occupies an angular slice of $\theta$.

We numerically optimized the bullseye structure using finite-difference time-domain simulations to achieve a resonance at 440 nm matching the impurity-bound exciton emission energy. From the simulations, we determined the optimal parameters to be D = 464 nm, $\Lambda$ = 116 nm, and $\Delta$ = 58 nm. With these parameters we attained a quality factor of 1100. From the simulations, we also found that the bridges can span an angular distribution of $\theta = 3°$ without degrading the cavity quality factor. Figure 1c shows a simulated on-resonance far-field radiation profile from the optimized optical cavity, with a nearly Gaussian distribution and ~37% of the radiation falling inside a 40-degree angle range defined by the 0.65 numerical aperture of the objective lens in our collection system. An associated near-field distribution is shown in Figure S3 (a). We provide complete details of the bullseye cavity design in Supplementary Materials Section II.

We fabricated the cavity using electron beam lithography and dry etching of the ZnSe quantum well membrane. First, we deposited a layer of SiN on top of the sample and then patterned it with the bullseye cavity design using electron beam lithography followed by inductively coupled plasma reactive ion etching using a $CHF_3/SF_6$ gas mixture. The SiN layer then served as a hard mask to etch the exposed ZnMgSe layer using inductively coupled plasma reactive ion etching with a $Cl_2/CH_4/H_2$ gas mixture. We then removed the SiN using the same etching process as before. Finally, we undercut the Zn(Mg)Se layer by wet-etching of the AlGaAs with hydrofluoric acid to create a suspended cavity. Fig. 1d shows the final bullseye cavity. A complete description of the fabrication process has been shown in Methods.

We first characterized the photoluminescence of the device in a closed-cycle refrigerator cooled down to 3.6 K. The sample was excited using an above-band continuous-wave laser at a wavelength of 405 nm and the emission was collected by an objective lens with a numerical aperture of 0.68. The photoluminescence measurements were performed in a grating spectrometer with a spectral resolution of 0.02 nm (128.25 $\mu$eV). Additional details of the experimental setup are provided in Methods.

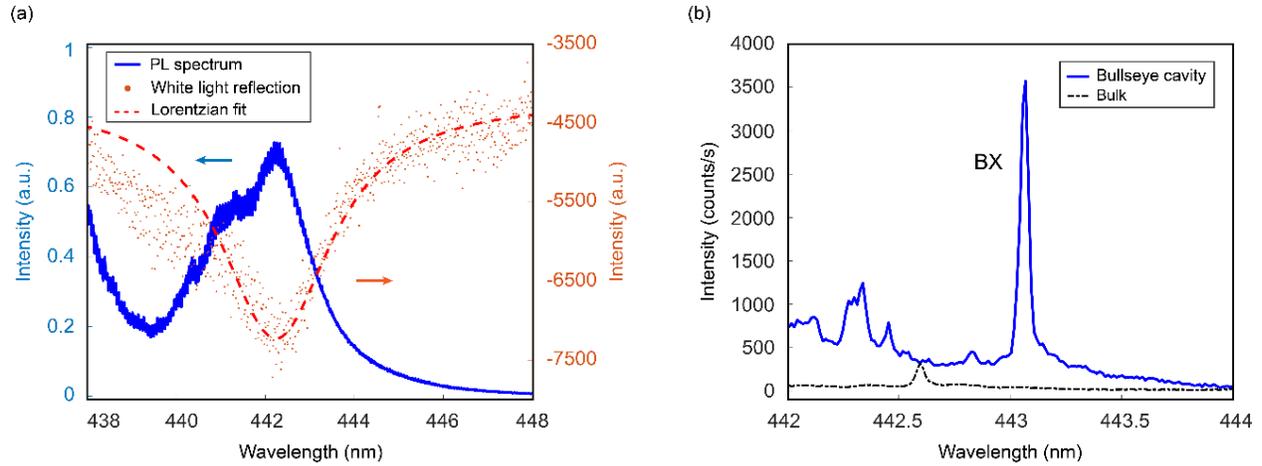

Figure 2. (a) Photoluminescence spectrum of a bullseye cavity device taken at an excitation power of 78 $\mu$W (blue solid line), along with the reflectivity spectrum obtained using white light excitation (red dots indicate data points and the dashed red line denotes a Lorentzian fit). (b) Photoluminescence spectrum from the same device obtained at an excitation power of 4 $\mu$W (blue line). The dashed black line shows the emission spectrum from the non-patterned, bulk region of the material at the same excitation power.

Figure 2a shows the photoluminescence spectrum obtained from the bullseye cavity using an above-band continuous wave laser at a wavelength of 405 nm and pump power of 78 $\mu$W. This excitation intensity saturates the quantum emitters, broadening their emission to form an internal wide-bandwidth light source that reveals a cavity mode at $\lambda_0 = 442.2$ nm (2.804 eV). We also measured the reflectivity spectrum (Figure 2a), which showed a reflectivity dip at the same wavelength as the photoluminescence emission peak, indicating this emission originates from the cavity mode. The cavity mode has a bandwidth of $\Delta\lambda = 0.926 \pm 0.010$ nm (5.889 $\pm$ 0.068 meV), which corresponds to a quality factor of $Q = \lambda_0/\Delta\lambda = 476$.

We next reduced the pump power to 4 $\mu$W to avoid saturating the emitter and suppress the charged exciton background emission. Figure 2b shows the resulting spectrum, which features a spectrally isolated impurity-bound exciton emission line at 443.064 nm, labeled "BX" that appears close to the wavelength of the cavity mode. The narrow linewidth of this bound exciton (0.040 $\pm$ 0.003 nm from fit, limited by the spectrometer resolution) suggests this peak originates from a single impurity-bound exciton emitter[12,13]. For comparison, we also measured the spectrum of an unpatterned region of the sample, as shown by the dashed line in Figure 2b, and observed a significantly dimmer emission from a different bound exciton. The brightness enhancement from the bullseye cavity region suggests a coupling between the impurity-bound exciton emitter and the optical cavity resonance.

To confirm the bound exciton line originates from a quantum emitter, we performed auto-correlation measurements using a pulsed laser at 405 nm wavelength, generated by frequency doubling a Ti:Sapphire laser. The pulsed excitation has a repetition rate of 76 MHz and a pulse duration of 2 ps. The sample was excited with an average pump power of ~1 $\mu$W and the emission was filtered by using a spectrometer grating and slit to isolate the bound-exciton line. The signal was measured by using an avalanche diode, followed by an electronic time interval analyzer that performs photon correlation measurements. Complete details of the experimental setup are provided in Methods.

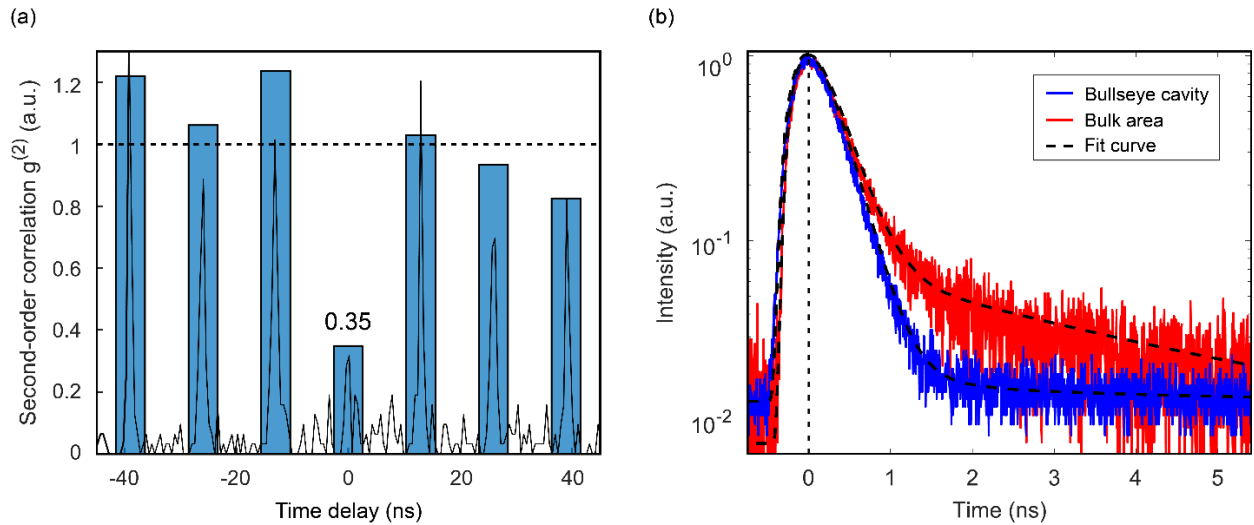

Figure 3. (a) Second-order correlation of the cavity-coupled bound-exciton emitter, where the bars plot the integrated photon counts for each pulse delay. The dip at zero pulse delay exhibits a reduced coincidence rate of $g^2(0) = 0.35$. (b) Time-resolved photoluminescence spectrum of the cavity-coupled bound-exciton emitter (blue line) under above-band excitation, along with a comparison measurement for a bulk emitter (red line). The dashed black lines denote biexponential fits of the decay curves.

Figure 3a shows the measured second-order correlation of a bound exciton emitter in the cavity. We plot both the raw coincidence data (black line) and the integrated counts over each laser pulse (blue bars). The accumulated counts are normalized to an average value of 10 adjacent pulse intensities, while the raw data is normalized to the average peak value of the adjacent laser pulses. Due to the bright emission of the bound exciton in the bullseye cavity, even without applying a background subtraction, we observe a clear second-order correlation dip at time delay $\tau = 0$, with $g^2(0) = 0.35$. This result indicates that the sharp bound-exciton emission peak originates from a single photon source. However, the optical cavity mode has a background emission that adds false counts to the correlation. To correct for these backgrounds, we calculate the background subtracted autocorrelation using the he equation $g_{corr}^2(\tau) = (g^2(\tau) - (1 - R^2))/R^2$, where $R = S/(S + B)$, S is the signal count rate from the quantum emitter, and B is the background count rate[29]. In our experiment, we obtain a ratio of $R = 0.86$, with signal count rate S is measured from the emission peak and background count rate B is measured by detuning the collecting filter window around 0.3 nm away from the emission line. From these values we calculate $g_{corr}^2(0) = 0.12$.

Figure 3b shows the time-resolved lifetime measurement of the bound-exciton emission line. The lifetime was measured using an avalanche photodiode with a temporal resolution of 50 ps. To attain better temporal resolution, we performed deconvolution of the lifetime response with the detector impulse response (EasyTau software, Picoharp). The bound-exciton line exhibits a biexponential decay with both a fast and slow decay component, consistent with previous measurements of Cl impurity-bound excitons[12]. The fast decay lifetime is $\tau_{C1} = 141.1 \pm 0.7$ ps and the slow decay is $\tau_{C2} = 0.617 \pm 0.062$ ns. For comparison, we measure the lifetime value of a bound-exciton emitter in the bulk, which exhibits a fast decay time of $\tau_{B1} = 201.6 \pm 5.9$ ps and a slow decay lifetime of $\tau_{B2} = 2.61 \pm 0.28$ ns, which is also consistent with previous measurement results[12]. Comparing the radiative decay rates of the cavity and bulk, we obtain a Purcell factor of $F_p = \gamma/\gamma_0 = \tau_{B1}/\tau_{C1} = 1.43 \pm 0.05$. We note this enhancement is significantly lower than the theoretically predicted value of around 24 based on the cavity quality factor (see Supplementary Materials Section II). We attribute this difference to possible imperfect alignment of the emitter with the high field region of the cavity, as well as imperfect dipole orientation. The reduced Purcell factor may also be indicative of the presence of non-radiative decay, which may be inherent to these emitters but may also be induced by surfaces introduced during the cavity fabricating process.

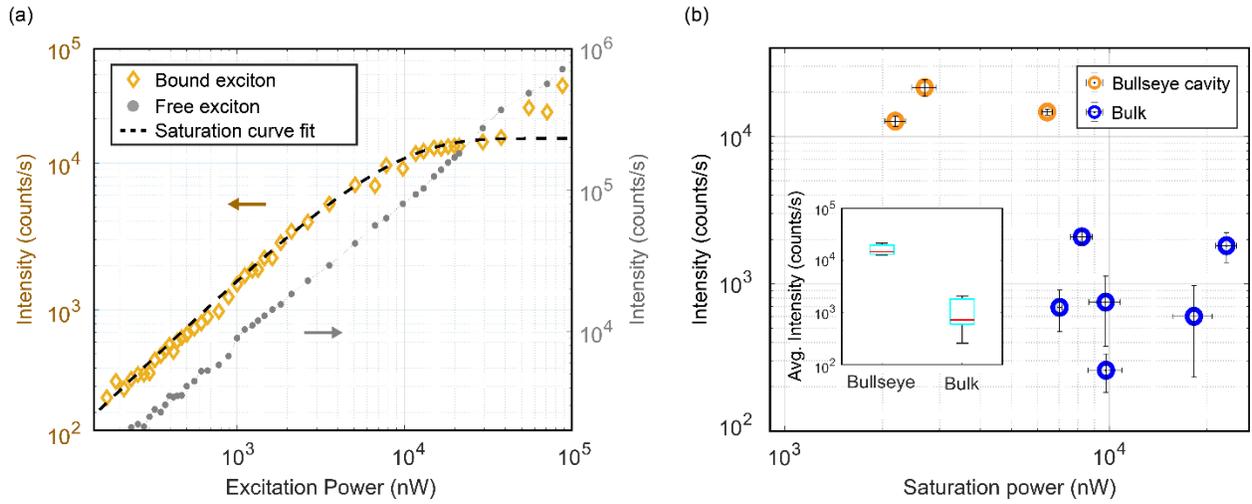

Figure. 4. Power-dependent photoluminescence intensity measurements. (a) Emission intensity from a bullseye cavity-coupled emitter as a function of the continuous-wave above-band excitation power. The photoluminescence intensity of a single bound exciton shows linear increase vs. the excitation power first and saturates as the excitation power increases towards high-power regime. The saturation behavior of a two-level system has been fitted using a numerical model. In contrast, a free-exciton emission does not show such saturation behavior. (b) We compare the saturation behavior of the bound-exciton emission in the bullseye cavity vs. the bulk for multiple emitters. (Inset) Comparison of the average intensity of the bullseye and bulk bound-exciton emitters. Cavity-enhancement increases the emission intensity by 15.8 times on average.

To characterize the brightness of the bound exciton quantum emitter, we measured the intensity of the emission line as a function of the excitation power (Figure 4a). The integrated intensity of the bound exciton (within a bandwidth of 0.091 nm) exhibits an initial linear increase and then saturates, consistent with the behavior of a two-level emitter. The dashed line shows a numerical fit of the experimental data to the equation $I = I_{sat} \exp(-P/P_0)^\alpha + I_0$ [21,30], where $I_{sat}$ is the saturation intensity of the emitter, $P_0$ is the threshold excitation power, $I_0$ is the background intensity, and $\alpha$ is a parameter that accounts for a linear correction[13]. From this fit, we determine the saturation intensity of this emitter to be $I_{sat} = 1.47 \pm 0.07 \times 10^4$ counts/s. For comparison we also plot the free-exciton excitation-power dependence in the same measurement, which increases linearly without saturation.

To quantify the brightness enhancement of the cavity, we perform the same measurement for multiple quantum emitters in both the bulk and bullseye cavities. Figure 4b shows the saturation intensities of multiple emitters from both types of emitters. The average saturation intensity of the bulk bound-exciton emitters is $I_{sat} = 1.03 \times 10^3$ counts/s while the average of the bound-exciton emitters in the cavity is $I_{sat} = 1.63 \times 10^4$ counts/s, constituting a 15.8-times enhancement in the emission brightness. This large enhancement demonstrates the effectiveness of the cavity at increasing the II-VI impurity-bound exciton emitter brightness via optical resonance.

**Conclusion**

In conclusion, we have demonstrated cavity-enhanced emission from a single impurity-bound exciton in a ZnSe quantum well. We observe a brightness enhancement of over an order of magnitude compared to bulk emitters, constituting a significant improvement. Further enhancement can be achieved by using cavity designs that exhibit higher quality factors[31,32], or smaller mode volumes[33], potentially enabling the strong light-matter regime. Impurity-bound excitons in ZnSe also exhibit a spin ground state[12], which could enable strong spin-photon interfaces mediated by the optical cavity. Ultimately, our results pave the way towards active nanophotonic systems where impurity-bound excitons can act as quantum light sources, strong optical nonlinear elements, and spin qubits with efficient photonic interfaces.

**Acknowledgement**

Support is acknowledged from AFOSR grant #FA95502010250 and The Maryland-ARL quantum partnership #W911NF1920181. This work is also funded by the Deutsche Forschungsgemeinschaft (DFG, German Research Foundation) under Germany's Excellence Strategy - Cluster of Excellence Matter and Light for Quantum Computing (ML4Q) EXC 2004/1−390534769.

**Additional information**

Supplementary information is available in the online version of the paper.

**Competing financial interests**

The authors declare no competing financial interests.

## Methods

*Device fabrication*

The ZnSe quantum well structure is grown by a molecular beam epitaxy process, as described in previous work[12,13]. To fabricate the bullseye cavity, we first grow a layer of 110 nm SiN using plasma-enhanced chemical vapor deposition with a deposition rate of 16.7 nm/min. The SiN is used to create a hard mask for the subsequent reactive ion etching processes on ZnSe. We spin-coat a thin layer of positive electron beam resist (diluted ZEP-520A, ZEP:Anisole = 5:3) at the spin rate of 4000 rpm with 1 min, which gives a thickness of around 120 nm. The sample is baked before and after spin coating at 100 ℃ for 2 minutes. We pattern the optimized bullseye cavity nanostructure design using electron beam lithography (Elionix ELS-G100). The SiN is etched through the ZEP electron beam resist mask after developing, by using inductively coupled plasma reactive ion etching with a $CHF_3/SF_6$ gas mixture. The etch rate was measured to be 4 nm/s. Then the Zn(Mg)Se layer is etched using the SiN layer as a mask by using inductively coupled plasma reactive ion etching with a $Cl_2/CH_4/H_2$ gas mixture. The SiN is removed by the same etching process as previously described. After patterning the nanostructure onto the Zn(Mg)Se layer, a wet etching process based on 0.5% hydrofluoric solution is used to etch the AlGaAs sacrificial layer. Using this concentration, an area of around $2\mu m \times 2\mu m$ of undercut is etched in 2 minutes.

*Experimental setup*

The sample is mounted in a 3.6K close-loop cryostat with an objective lens which has a numerical aperture of 0.68 and a focal length of 3 mm. The sample could be excited using either a 405 nm continuous wave laser diode (Thorlabs) or a frequency doubled pulsed Ti:Sapphire laser (Mira). The collected photoluminescence was measured spectrally using a spectrometer (Princeton Instrument) with a grating of 1800 g/mm and a 1340 pixels CCD detector to provide a spectral resolution of 0.02 nm. The second order correlation was measured by a Hanbury-Brown and Twiss setup by splitting the beam into two paths and detecting photons using avalanche photodiodes (Micro Photon Devices). A time interval analyzer was used to analyze the intensity versus the time delay (Time-correlated Single Photon Counter, Picoharp 300). The pulsed laser was tuned to 405 nm for above band excitation to acquire the second-order correlation function and the time-resolved photoluminescence spectrum.